\documentclass[runningheads]{llncs}

\usepackage[T1]{fontenc}
\usepackage{graphicx}
\usepackage{todonotes}
\usepackage{enumitem}
\usepackage{booktabs}
\usepackage{longtable}
\usepackage{makecell}

\usepackage{amssymb}
\usepackage{todonotes}
\usepackage{hyperref}
\usepackage{comment}
\usepackage{booktabs}
\usepackage{pdflscape}
\usepackage{graphicx}
\usepackage{textcomp}
\usepackage{multicol}
\usepackage{tabularx}
\newcommand\setrow[1]{\gdef\rowmac{#1}#1\ignorespaces}
\newcommand\clearrow{\global\let\rowmac\relax}
\clearrow

\usepackage{outlines}
\usepackage{multirow}
\usepackage{lscape}
\usepackage{comment}
\usepackage[table,xcdraw]{xcolor}
\usepackage{float}
\usepackage{enumitem}
\usepackage{ctable}
\usepackage{caption}
\usepackage[many]{tcolorbox}
\usepackage{nameref}
\usepackage{hyperref}
\usepackage{makecell}
\usepackage{hhline}
\usepackage{ctable}
\usepackage{array}
\usepackage{comment}
\usepackage{longtable}
\usepackage[font=small,skip=-2pt]{caption}
\usepackage[T1]{fontenc}
\usepackage{array}

\usepackage{lmodern}

\newtcolorbox{new}[1][]{
colback  = white,
coltitle = white,  
outer arc=0pt,
arc=0pt,
left=0pt,
right=0pt,
top=0pt,
bottom=0pt,
title    = {\bf Bảng ~\thetcbcounter} ,
#1,
}

\newtcolorbox{colorb}{
enhanced,
boxrule=0pt,
frame hidden,
borderline west={2pt}{0pt}{green!50!black},
colback=green!05!white,
sharp corners
}

\newtcolorbox{colora}{
enhanced,
top=1pt,
bottom=1pt,
right=1pt,
left=4pt,
boxrule=0pt,frame hidden,
borderline west={2pt}{0pt}{gray!50!black},
colback=gray!05!white,
sharp corners
}

\begin{document}

\title{Towards a Goal-Centric Assessment of Requirements Engineering Methods for Privacy by Design}

\titlerunning{Towards a Goal-Centric Assessment of RE Methods for PbD}

\author{Oleksandr Kosenkov\inst{1,2} \and
Ehsan Zabardast\inst{2} \and
Jannik Fischbach\inst{1}
\and
Tony Gorschek\inst{1,2} \and Daniel Mendez\inst{1,2}
}

\authorrunning{O. Kosenkov et al.}

\institute{fortiss GmbH, Munich, Germany \\
\email{lastname@fortiss.org}
\and
Blekinge Institute of Technology, Karlskrona, Sweden \\
\email{firstname.lastname@bth.se}
\\}

\maketitle

\vspace{-20pt}
\begin{abstract}
Implementing privacy by design (PbD) according to the General Data Protection Regulation (GDPR) is met with a growing number of requirements engineering (RE) approaches. However, the question of which RE method for PbD fits best the goals of organisations remains a challenge. We report our endeavor to close this gap by synthesizing a goal-centric approach for PbD methods assessment. We used literature review, interviews, and validation with practitioners to achieve the goal of our study. As practitioners do not approach PbD systematically, we suggest that RE methods for PbD should be assessed against organisational goals, rather than process characteristics only. We hope that, when further developed, the goal-centric approach could support the development, selection, and tailoring of RE practices for PbD.

\end{abstract}

\section{Introduction}
Achieving GDPR compliance during software engineering (SE), especially in the requirements engineering (RE) phase of the software development life cycle (SDLC). The number of RE methods trying to address  GDPR compliance is constantly growing. In practice, the adoption of these methods and tools remains poorly supported~\cite{klymenko2022understanding,alhirabi2023parrot}, and ad-hoc approaches are widespread. One reason for this is the absence of concrete assessment criteria~\cite{alhirabi2023parrot} for RE methods. The adoption of existing contributions in practice becomes cumbersome
without any guidance on the selection and implementation of RE approaches. Existing SE process assessment approaches fail to address the specificity required for GDPR compliance. First, GDPR interpretation is complex and should address legal goals in a valid way in variable organizational conditions. Second, the GDPR mandates privacy by design (PbD), requiring regulatory requirements to be implemented throughout SDLC via RE and, in particular, integrated into requirements and system specification artefacts. Still, existing approaches consider requirements and system specifications in isolation. Finally, regulatory compliance is an enterprise-level concern~\cite{kosenkov2024regulatory} requiring coordination and assessment of PbD methods across multiple development teams towards the achievement of enterprise goals. Yet, existing approaches do not sufficiently account for such goals. To facilitate the assessment of RE methods for PbD, we conducted a literature review, interviews, and synthesized and validated an initial version of a goal-centric assessment approach. Our approach is intended to guide organizations in choosing and tailoring or developing RE methods for PbD. We present our preliminary results to foster discussion and invite community feedback on the idea of goal-centric assessment of RE methods.

\vspace{-0.15cm}
\section{Background}\label{sec:background}
\vspace{-0.15cm}
Both ``personal data protection by design'' (Art. 25 GDPR) and the PbD principle, formulated in research, require controls to be considered from the beginning of the SDLC and implemented in the software design. We understand \textit{privacy by design (PbD)} as the specification and implementation of requirements and system architecture in response to GDPR norms. We define \textit{requirements engineering (RE) for PbD} as the requirements specification that captures requirements and early architecture (constraints) in response to GDPR norms, and effectively and correctly transmits specifications for implementation in the software design. We define \textit{RE method characteristic} as a quality of an outcome and RE artifacts obtained by applying a method (e.g., specifications consistency). \textit{RE method goal} is an ultimate purpose for which a method outcome is used and which drives the RE process (e.g., consistency supports the goal of managing risks).

\vspace{-0.15cm}
\section{Related Work}\label{sec:relatedWork}
\vspace{-0.15cm}
Existing studies suggest multiple RE methods for PbD or GDPR compliance (e.g., \cite{kosenkov2024systematic} identified 15 studies covering the RE-software design intersection for GDPR compliance), however, only some conduct an evaluation in practical settings or apply PbD-specific criteria. Studies in business process management and legal informatics suggest some systematic evaluation approaches for GDPR text processing (e.g., \cite{bartolini2018agile}), but do not consider such approaches in the context of SE.
There are some general frameworks and approaches for SE and RE process assessment and improvement, such as CMMI, SPICE / ISO 33061, iFLAP, and Uni-REPM framework. However, these approaches do not address regulatory compliance, lack problem-orientation, explicit statement of method goals and improvements, and rather focus on process and/or product characteristics~\cite{unterkalmsteiner2011evaluation}.

\vspace{-0.15cm}
\section{Methodology}\label{sec:methodology}
\vspace{-0.15cm}
Next, we report the methodology applied in this study to the extent necessary to explain the development of our goal-centric assessment approach and supports its subsequent discussion (for details \href{https://zenodo.org/records/15760786}{see open data set} and Fig.~\ref{fig:methodology} for visual overview). This study was guided by the following research questions:

\begin{figure}[ht!]
    \centering
    \includegraphics[width=1\linewidth]{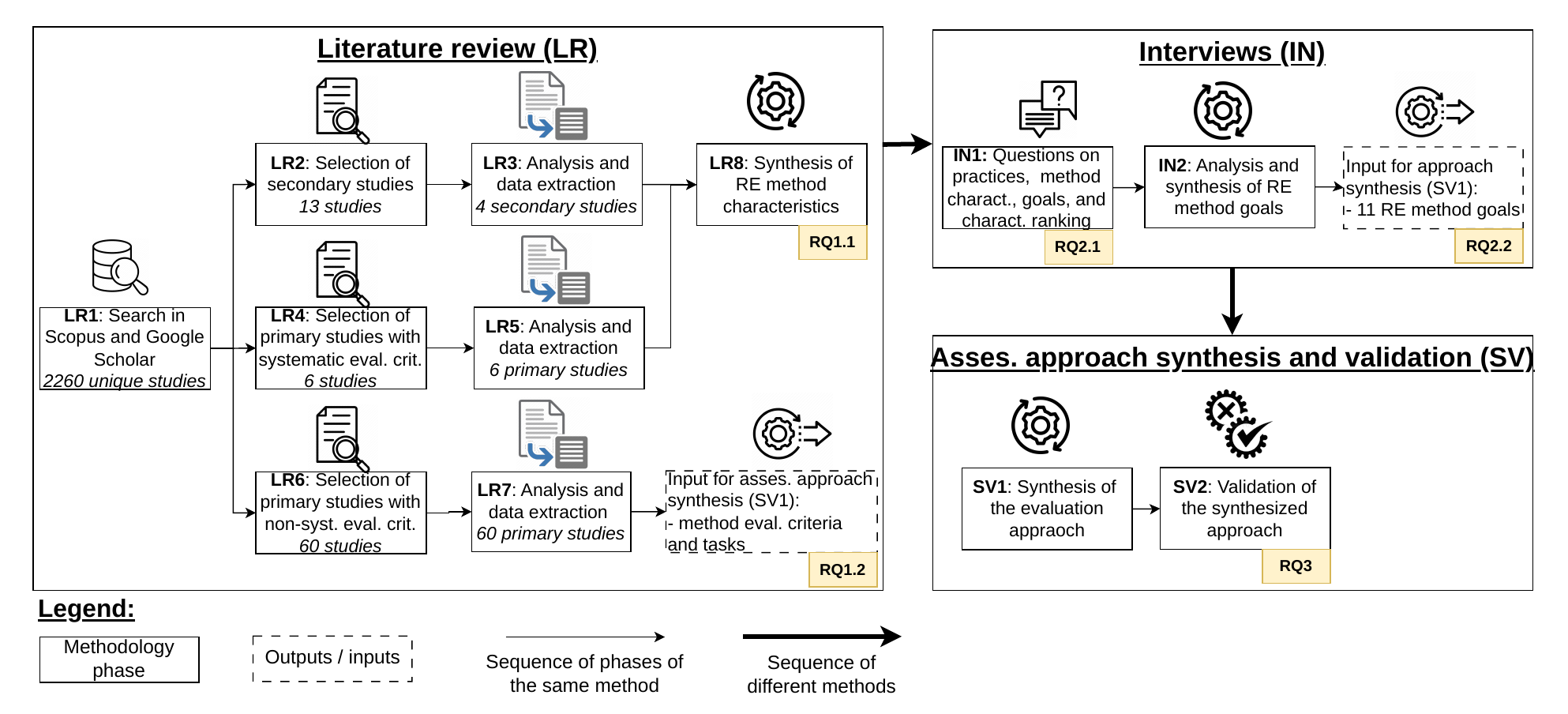}
    \caption{Visual overview of the methodology}\label{fig:methodology}
\end{figure}

\begin{itemize}[nosep, leftmargin=1.25cm]
    \item[RQ1.1:] What are the reported evaluation criteria for methods for PbD?
    \item[RQ1.2:] What are the reported characteristics of methods for PbD?
    \item[RQ2.1:] How do practitioners assess the method characteristics importance?
    \item[RQ2.2:] What are the goals of practitioners in the process of RE for PbD?
    \item[RQ3:] How useful and feasible is a goal-centric approach to the assessment of RE methods for PbD, according to practitioners?
\end{itemize}

To answer RQ1.1, RQ1.2 we executed a literature review and synthesized five RE method characteristics (MCs) essential for PbD. We conducted interviews with practitioners to answer RQ2.1, RQ2.2, and identified eleven method goals (MGs) to which method characteristics contribute. Finally, we answered RQ3 by validating the initial version of the assessment approach.

\textit{Literature Review (LR)}. After trial searches with different keywords, we excluded potentially limiting auxiliary terms (e.g., compliance) and executed the following query in Scopus and Google Scholar (without filters): \textit{gdpr AND ( ``software engineering'' OR ``requirements engineering'' ) AND ( evaluation OR validation )}. We retrieved \textbf{2260} unique primary studies (1777 from Scopus, the first 700 results from Google Scholar, and after removing 217 duplicates). We have not found any approach for systematically assessing RE methods for PbD, and, thus, we extracted the data for its synthesis from existing secondary studies (LR2-3), primary studies suggesting and evaluating new PbD methods systematically with tailored criteria (LR4-5), and primary studies suggesting new PbD methods, but assessing them with non-tailored criteria (LR6-7). We also extracted criteria from three publications on systematic evaluation of RE methods for system specification (e.g.,~\cite{galster2009comparing}). We applied a thematic analysis~\cite{braun2006using} and meta synthesis~\cite{sandelowski2006handbook} to synthesize (1) the RE method characteristics (LR8) using the data on systematic assessment of RE methods for PbD (LR2-5, and additional publications), and (2) our evaluation approach (SV1) using the data about non-systematic evaluation (LR6-7) in addition to the interviews results (IN1-2).

\textit{Interviews (IN)}. To answer RQ2.1, RQ2.2 we conducted semi-structured interviews~\cite{runeson2009guidelines}. We applied purposive sampling with snowballing to select participants involved in RE for PbD/GDPR compliance, with both technical insights and experience with GDPR, including both engineering roles (e.g., architects) and roles collaborating with them (e.g., lawyers).
We adapted the Goal Question Metric (GQM) approach~\cite{van2002goal} to structure the interview process and further analysis. The GQM is an approach to defining and measuring software quality at three levels: (1) conceptual level, defining goals to be achieved, (2) operational level, defining questions for evaluating the achievement of goals, and (3) quantitative level, defining metrics or data required to answer the questions. For each method characteristic, we formulated open-ended questions about goals, questions, and metrics. At the end of the interviews, we asked to rank MCs (from 1---most important to 5---least important).

\textit{Synthesis \& Validation of the Assessment Approach (SV)}.
We synthesized the initial version of the assessment approach on the basis of the GQM approach, as briefly described above. The RE method goals identified in the interviews (IN2) served as a pivot for this. To iteratively develop our approach, we have used (1) questions, metrics, and additional comments identified in the interviews, (2) the relationship between method characteristics and goals, (3) the ranking of method characteristics assigned during the interviews, (4) evaluation criteria and tasks extracted from the literature. Where appropriate, we further decomposed method goals into subgoals. Lastly, the first author refined the approach on the basis of his knowledge of GDPR. We validated the usefulness and feasibility of our approach using screening and walkthrough of the assessment approach~\cite{kitchenham1996desmet} (see Tab.~\ref{tab:example} for an example of the approach provided to participants). Validation participants evaluated the \textit{usefulness} (a quality of the components of our approach of being useful) and \textit{feasibility} (a quality of being reasonable and likely to be applied for RE method assessment in practice) of each element of the approach (subgoals, questions, metrics) as positive or negative. See Tab.~\ref{tab:interviewResults} for the number of positively evaluated components in relation to the total number of corresponding components (with the remaining evaluations being negative).

\vspace{-0.15cm}
\section{Results}\label{sec:results}
\vspace{-0.15cm}
\paragraph{Literature Review Results}
The intersection between requirements and system specification for GDPR compliance is mentioned in multiple studies; however, there are no systematic approaches for the assessment of methods for PbD in practical settings. We found 6 primary studies suggesting new methods for PbD, which recognized and used a tailored evaluation of their contributions. The majority of the other primary studies have reused existing non-PbD-specific criteria for evaluation of their contributions. In the literature review, we extracted the following relevant data: (1) tasks, stakeholders, challenges to GDPR compliance from 4 relevant secondary studies (e.g.,~\cite{kosenkov2024systematic}, (2) evaluation criteria applied for systematic evaluation of PbD methods in 6 primary studies (e.g.,~\cite{bartolini2018agile}), and (3) criteria and tasks (e.g., requirements conflict identification) for non-systematic evaluation of newly suggested PbD methods from 60 secondary studies.

\begin{colora}
\small
\textbf{RQ1.1:}
\textit{The core reported systematic evaluation criteria for PbD methods are correctness, transparency, support of legal goals, activities, and documentation, support for both architectural and legal concepts.}
\end{colora}

\begin{colora}
\small
\textbf{RQ1.2:}
\textit{The five basic characteristics of RE methods (MCs) for PbD are as follows:}
\end{colora}

\textit{MC1: Capturing legal domain knowledge} is fundamental for represent legal concepts, identifying goals and architecturally significant requirements, and resolving the GDPR abstractness and engineering-legal perspectives gap.
\textit{MC2: Traceability \& Consistency of Specifications} is mainly important to address challenges of certification and compliance provability.
\textit{MC3: Separation of Compliance \& Non-Compliance Concerns} is important as regulated system components can require specific handling, and compliance can conflict with existing SE methods (e.g., outsourcing).
\textit{MC4: System Specification Transparency} enables involvement of stakeholders and supports their activities (e.g., risk management).
\textit{MC5: System Specification Flexibility} facilitates the architecture flexibility in response to changes in regulations or software, and maintenance of corresponding models.

\vspace{-0.5cm}
\begin{table}[ht!]
\tiny
\centering
\rowcolors{4}{gray!15}{white}
\begin{tabular}{cclcc|ccccc|cc|cc|ccc}

&  &  &  &  & \multicolumn{5}{c|}{\tiny{Ranking}} & \multicolumn{2}{c|}{\tiny{\parbox{1cm}{Subg. \\ (of 32)}}} &  \multicolumn{2}{c|}{\tiny{\parbox{1cm}{Quest. \\(of 148)}}} & \multicolumn{2}{c}{\tiny{\parbox{1cm}{Metric \\(of 172)}}}  & \\

\multicolumn{1}{c}{\thead{{\tiny{ID}}}} & \multicolumn{1}{c}{\thead{\tiny{Comp.}}} &  \multicolumn{1}{c}{\thead{\tiny{Role}}} & \thead{\tiny{Exp.}} & \thead{\tiny{GDPR Exp.}} & \thead{\tiny{MC1}} & \thead{\tiny{MC2}} & \thead{\tiny{MC3}} & \thead{\tiny{MC4}} & \thead{\tiny{MC5}} & U & F & U & F & U & F \\

\hline
I1 & C1 & Tech lead & 4 & 3 & 5 & 4 & 2 & 3 & 1 & - & - & - & - & - & - \\ 
I2 & C1 & Sales engineer & 4 & 6 & 4 & 1 & 5 & 2 & 3 & - & - & - & - & - & - \\ 
I3 & C1 & Stream lead & 28 & 6 & 3 & 4 & 1 & 2 & 5 & - & - & - & - & - & - \\ 
I4 & C1 & Data engineer & 3,5 & 3,5 & 1 & 2 & 5 & 3 & 4 & - & - & - & - & - & - \\
I5 & C2 & Web marketing spec. & 8 & 5 & 1 & 2 & 5 & 4 & 3 & 32 & 32 & 146 & 137 & 167 & 158 \\
I6 & C3 & Security manager & 34 & 7 & 1 & 2 & 5 & 4 & 3 & - & - & - & - & - & - \\
I7 & C4 & Data engineer & 4 & 2 & 2 & 1 & 5 & 4 & 3 & - & - & - & - & - & - \\ 
I8 & C5 & IT Compl.\&Audit Head & 30 & 7 & 2 & 3 & 1 & 5 & 4 & - & - & - & - & - & - \\ 
I9 & C6 & Cloud Sec. Architect & 14 & 5 & 1 & 3 & 5 & 2 & 4 & - & - & - & - & - & - \\
I10 & C7 & Architect \& Req-s Eng. & 7 & 6 & 1 & 2 & 5 & 3 & 4 & - & - & - & - & - & - \\
I11 & C8 & Project manager & 20 & 7 & 1 & 2 & 3 & 4 & 5 & - & - & - & - & - & - \\
I12 & C8 & Tech. project manager & 3 & 3 & 1 & 3 & 4 & 2 & 5 & - & - & - & - & - & - \\
I13 & C9 & Software developer & 12 & 7 & 1 & 4 & 5 & 3 & 2 & 32 & 27 & 146 & 132 & 170 & 152 \\
I14 & C10 & Software architect & 15 & 7 & 1 & 4 & 5 & 3 & 2 & - & - & - & - & - & - \\
I15 & C11 & Info. Security Expert & 3 & 8 & 1 & 4 & 5 & 3 & 2 & - & - & - & - & - & - \\
I16 & C12 & Data Protection Officer & 7 & 7 & - & - & - & - & - & 32 & 32 & 147 & 141 & 152 & 142 \\
I17 & C13 & Data Protection Advisor & 4 & 4 & - & - & - & - & - & 31 & 16 & 137 & 98 & 158 & 75 \\
I18 & C14 & Cloud Sec. Architect & 2.5 & 5 & - & - & - & - & - & 31 & 30 & 140 & 131 & 150 & 141 \\
I19 & C15 & Software Developer & 7 & 2 & - & - & - & - & - & 29 & 28 & 141 & 135 & 159 & 152 \\
\hline

\multicolumn{2}{l}{\textbf{Median}} &  & \textbf{} & \textbf{} & \textbf{1} & \textbf{3} & \textbf{5} & \textbf{3} & \textbf{3} & NA & NA & NA & NA & NA & NA & \rowcolors{0}{white}{white}\\
\multicolumn{2}{l}{\textbf{Mode}} &  & \textbf{} & \textbf{} & \textbf{1} & \textbf{2, 4} & \textbf{5} & \textbf{3} & \textbf{3, 4} & NA & NA & NA & NA & NA & NA & \rowcolors{0}{white}{white}\\
\multicolumn{2}{l}{\textbf{Sum}} &  & \textbf{} & \textbf{} & \textbf{26} & \textbf{41} & \textbf{61} & \textbf{47} & \textbf{50} & NA & NA & NA & NA & NA & NA & \rowcolors{0}{white}{white}\\
\end{tabular}
\caption{Overview of interviewee ID, company ID, role, general experience, experience with GDPR, ranking of method characteristics and evaluation of usefulness (U) and feasibility (F) of assessment approach components (subgoals, questions, metrics).}
\label{tab:interviewResults}
\end{table}

\vspace{-1.2cm}

\paragraph{Interview Results}
Only 2 out of 15 interviewees used a specific RE method for PbD, I8 used a tool for compliance in different jurisdictions, and I14 used a data protection impact assessment tool for collaboration with lawyers. Other interviewees used ad hoc approaches (e.g., I1 interviewed lawyers). I1, I11, I14 reported having used some ($\leq$4) of the questions and metrics in practice; the remainder were not previously used but potentially applicable. The highest number of the 190 goals was identified in connection with MC4 specification transparency (45 goals), and MC1 capturing legal knowledge (39 goals) (see Tab.~\ref{tab:mc-mg}).

\begin{colora}
\small
\textbf{RQ2.1:}
\textit{The ranking of the RE method characteristics according to their importance is as follows: MC1: capturing legal knowledge, MC2: traceability and consistency, MC4: specification transparency, MC5: system specification flexibility, MC3: separation of compliance and non-compliance concerns (see Tab.~\ref{tab:interviewResults}.}
\end{colora}

\begin{colora}
\small
\textbf{RQ2.2:}
\textit{After a thematic analysis of the goals, we synthesized the following eleven core RE method goals for PbD (MGs):}
\end{colora}

\textit{MG1: Facilitating GDPR compliance throughout SDLC} includes subgoals for facilitating compliance in the design (e.g., controls modularity), implementation, and testing phases of SDLC, and integration into SDLC models. 

\textit{MG2: achieving understandability of GDPR} subgoals are related to understandability for the involved roles,  common understanding, awareness about related aspects (e.g., risks), clarity of GDPR-to-system mapping, sustaining the understanding, best practices synthesis, and understandability of automation.

\textit{MG3: enabling decision-making} includes subgoals for enabling decisions, compliance efficiency, balancing compliance, technical, business, and other needs, planning expenses, and effectively managing non-compliance (e.g., financial risk).

\textit{MG4: Documenting the required information} includes subgoals addressing documentation availability, versioning, and history, documentation content, and characteristics, and documentation of GDPR-to-system mapping.

\textit{MG5: Facilitating compliance governance} includes subgoals on using the processes supporting compliance (e.g., change management), identification of the required information, compliance gaps and the overall compliance status, execution of other governance types (e.g., IT governance), and compliance manageability.

\textit{MG6: Enabling response to changes} subgoals focused on change identification (e.g., identifying what is likely to change), and facilitating the implementation of changes (e.g., implementation without impacting other components).

\textit{MG7: Verifiability\&Validity of compliance} has subgoals for clarity and availability of information for verification, capacity to trace the required information, addressing validity from stakeholders' perspectives, clarity of the information about the perspectives, sustaining validity, and knowing barriers to it.

\textit{MG8: Achieving effective communication on GDPR compliance implementation} subgoals under MG consider the facilitation of cross-functional communication, achieving interaction between roles, translating between legal, technical, and business requirements, and communication in a common language.

\textit{MG9: Facilitating GDPR compliance-related procedures} focuses on facilitating procedures such as audits and compliance reviews.

\textit{MG10: Addressing business concerns in PbD implementation} is connected to facilitating business processes improvement, maximization of business outcomes.

\textit{MG11: Facilitating risk and security management} focuses on intersections between GDPR compliance and security risks, controls, and their management.

\begin{table}[ht!]
\scriptsize
\centering
\rowcolors{2}{gray!15}{white}
\begin{tabular}{|>{\rowmac}c|>{\rowmac}c|>{\rowmac}c|>{\rowmac}c|>{\rowmac}c|>{\rowmac}c|>{\rowmac}c|>{\rowmac}c|>{\rowmac}c|>{\rowmac}c|>{\rowmac}c|>{\rowmac}c||>{\rowmac}>{\bfseries}c|}
\hline
\thead{{\scriptsize{MC}}} & \thead{\scriptsize{MG1}} &  \thead{\scriptsize{MG2}} & \thead{\scriptsize{MG3}} & \thead{\scriptsize{MG4}} & \thead{\scriptsize{MG5}} & \thead{\scriptsize{MG6}} & \thead{\scriptsize{MG7}} & \thead{\scriptsize{MG8}} & \thead{\scriptsize{MG9}} & \thead{\scriptsize{MG10}} &  \thead{\scriptsize{MG11}} & \thead{\scriptsize{Sum}} \\
\hline
MC1 & 3 & 14 & 6 & 4 & 2 & 1 & 2 & 4 & - & - & 3 & 39 \\
MC2 & 7 & 3 & 3 & 6 & 4 & 4 & 5 & - & 1 & - & - & 33 \\
MC3 & 4 & 2 & 6 & 2 & 2 & 1 & 3 & 2 & 1 & 1 & - & 24 \\
MC4 & 4 & 8 & 7 & 10 & - & 1 & 1 & 5 & 4 & 1 & 4 & 45\\
MC5 & 9 & 1 & 3 & 1 & 4 & 10 & - & 1 & 1 & - & - & 30\\
MC6 & 3 & 2 & 1 & 1 & 7 & - & - & 1 & 1 & 3 &  & 19 \\
\hline\hline
\setrow{\bfseries}Sum & 30 & 30 & 26 & 24 & 19 & 17 & 11 & 13 & 8 & 5 & 7 & 190\\
\hline
\end{tabular}
\caption{Overview of the number of method goals (MGs) mentioned in connection to method characteristics (MCs).}
\label{tab:mc-mg}
\end{table}

Table~\ref{tab:mc-mg} shows the number of goals grouped by MGs we synthesized and in relation to MCs with respect to which goals were mentioned. This data indicates that in the future it will be important to identify the contribution of MCs to MGs (e.g., MC1 capturing legal knowledge to MG2 achieving understandability).

\paragraph{Synhtesized Approach}
The initial version of the approach was elaborated only to a limited degree, allowing for the validation of the core idea of building the RE method assessment around the method goals, rather than process characteristics, and starting a discussion with the community. The approach (see Tab.~\ref{tab:example} for an excerpt), similarly to GQM, should allow the goal-centric assessment of RE methods for PbD (incl. ad hoc methods) on three levels of abstraction (1) a conceptual with 11 goals / 32 subgoals predefined (see~\ref{sec:results}) that should be achieved in the process of RE for PbD, (2) an operational (148 questions) focusing on the areas requiring assessment for the goals achievement, and (3) a quantitative with 172 metrics defining concrete criteria and metrics for addressing the corresponding questions. Unlike the GQM, our suggested assessment approach predefines the core approach components to embed the best practices for PbD. The approach can enable flexible assessment by allowing (1) selecting the relevant goals/subgoals requiring assessment and thereby scoping the corresponding questions and metrics, (2) extending existing components with the new ones (which will require development of the guidance for approach extension in the future; to date, we have not received any feedback about additional components required).

\vspace{-0.5cm}
\begin{table}[H]
\tiny
\centering
\begin{tabular}{|p{2cm}|p{1.5cm}|p{5cm}|p{3.2cm}|}
\hline
\textbf{Goals} & \textbf{Subgoals}
&  \textbf{Questions} & \textbf{Criteria/Metrics} \\ \hline

\multirow{4}{2cm}{G1: Facilitating GDPR implementation throughout the SDLC} & \multirow{4}{1.5cm}{G1.1: facilitating software design and architecture} & Q1.1.1: Does method support documenting GDPR compliance on software architecture level? & M1.1.1.1: documentation completeness \\
\cline{3-4}
 &  & Q1.1.2: Does method support architecture review engaging legal experts?  & M1.1.2.1: reviews frequency \\
\cline{3-4}
 &  & Q1.1.3: Is it possible to discern GDPR compliance controls that need to be implemented? & M1.1.3.1: number of controls identified \\
\cline{3-4}
 &  & Q1.1.4: Is it clear what is the priority for controls implementation? & M1.1.4.1: prioritization available \\
\hline

\end{tabular}
\caption{An excerpt from the evaluation framework illustrating the structure of the framework}\label{tab:example}
\end{table}
\vspace{-1cm}

The approach is applied in two phases (1) tailoring and (2) assessment. Following steps are required for tailoring (1.1) person conducting the assessment (assessor) reviews and selects the goals that organization is aiming to achieve in PbD and which require assessment, (1.2) subgoals belonging to selected goals are reviewed and selected, (1.3) questions belonging to selected subgoals are reviewed and selected, stakeholders required to answer these questions are identified, (1.4) the assessor independently or together with involved stakeholders selects the relevant metrics, and specifies the concrete steps for data collection (e.g., sources of data). This tailoring phase enables scoping of the assessment according to organizational needs. For the assessment execution, (2.1) data is collected, (2.2) answers to the questions are documented on the basis of the collected data, (2.3) subgoals achievement is documented, (2.4) the goals achievement and overall assessment results are documented. For the validation, only a general description and overview of the approach components (as in Tab.~\ref{tab:example} and the open data set) were provided without the guidance for application.

\paragraph{Validation Results} During the validation, participants found the idea and structure of the approach to be clear, and mainly, questions related to concrete components emerged. Overall, the approach was evaluated positively (see next minimal and maximal number of components considered useful or feasible across validation participants or Table~\ref{tab:interviewResults} for details). Reported non-feasibility of certain components was primarily related to participants' belief that specific concerns cannot be effectively addressed (e.g., communication in a common language).

\begin{colora}
\small
\textbf{RQ3:}
\textit{Min 29 (90\%)---max 32 (100\%) of 32 subgoals were evaluated as useful, 16 (50\%)---32 (100\%) subgoals were considered feasible. 137 (92\%)---147 (99\%) of 148 questions evaluated useful and 98 (66\%)---141 (95\%) feasible. 150 (87\%)---170 (99\%) of 172 metrics were considered useful and 75 (44\%)---158 (92\%) feasible.}
\end{colora}

\vspace{-0.15cm}
\section{Discussion}\label{sec:discussion}
\vspace{-0.15cm}
The conceptualization of PbD for GDPR compliance as a conjoint requirements and system specification was well received and aligned with the experience of practitioners. Our results point to the low maturity of RE methods for PbD in practice, for example, interviewees struggled to articulate their practices, some interviewees could not make an assessment, or changed it while answering. We suggest that such difficulties stemmed, at least in part, from the complex relations between method characteristics and goals. The positive validation of the initial version of our goal-centric approach provides the first indication that goals may be more suitable for assessment purposes. Our study shows that capturing legal knowledge (MC1) and specifications transparency (MC4) are critical for PbD, along with traceability (MC2). Also, practitioners need RE methods that not only support the subsequent SDLC phases and implementation of compliance controls but also account for internal organizational stakeholders (e.g., legal experts) and support their goals and activities (e.g., risk management). Our results support and concretize the idea that RE plays a fundamental role in supporting other phases of SDLC and is intertwined with them~\cite{nuseibeh2000requirements}, and further suggest that RE contributes to the fulfillment of organizational goals and activities.

\vspace{-0.15cm}
\section{Limitations \& Threats to Validity}\label{sec:threats}
\vspace{-0.15cm}
To mitigate threats to the validity of the results, we followed guidelines for literature review~\cite{kitchenham2007guidelines}, thematic analysis~\cite{braun2006using}, interviews~\cite{runeson2009guidelines}, and evaluation~\cite{kitchenham1996desmet}.
In the literature review, the authors used the selection criteria that did not require interpretation (e.g., defined evaluation criteria) and jointly discussed the intermediary results.
We applied interview questions formulated and scoped using the structured GQM approach and asked about any additional considerations.
Simplistic binary evaluation of the approach during the validation mainly served the purposes of the initial validation of the goal-centric assessment idea and identification of further improvement directions.
To partially mitigate the threats to the results' generalizability, we involved participants in different professional roles, engaged in software requirements and system specification for GDPR compliance, and had expertise in both GDPR and software technologies.

\vspace{-0.25cm}
\section{Conclusion}\label{sec:conclusion}
\vspace{-0.15cm}
Assessing RE methods for software compliance is essential due to their specificity, yet it is not systematic in practice. As RE for PbD demands the coordination of requirements with early architecture and subsequent SDLC phases, RE methods must capture legal knowledge and facilitate transparency of specification, along with providing traceability. Although practitioners often struggle to reason about practices in terms of method characteristics, they frequently pursue the same goals via different method characteristics, motivating a goal-centric assessment of RE methods. The validation of our vision of a GQM-inspired operationalization of such assessment points to its potential usefulness and feasibility in practice. Before further developing the approach, we invite community feedback on the goal-centric RE method assessment and its synthesis process.

\noindent The open data is hosted on \href{https://zenodo.org/records/15760786}{Zenodo (10.5281/zenodo.15760786)} and \href{https://regulatory-re.com/refsq2026/}{website}.
\vspace{-0.15cm}

\bibliographystyle{splncs04}
\bibliography{bibliography}
\end{document}